\documentclass[showpacs, aps,prd,twocolumn,superscriptaddress, preprintnumbers]{revtex4-1}

\usepackage{units}
\usepackage{upgreek}
\usepackage{amsmath,amssymb}
\usepackage{lineno}
\usepackage{subscript}
\usepackage{graphicx}
\usepackage{subfigure} 
\usepackage{epsfig}
\usepackage{times}

\usepackage{xcolor}
\definecolor{darkred}{rgb}{0.5,0,0}
\definecolor{darkblue}{rgb}{0,0,0.5}
\definecolor{firebrick}{rgb}{0.75,0.125,0.125}
\definecolor{darkgreen}{rgb}{0,0.5,0}
\usepackage[colorlinks=true,linkcolor=firebrick,citecolor=darkgreen,urlcolor=darkblue]{hyperref} 

\usepackage{xspace}
\def\degree{$^\circ$\xspace}
\def\Offline{\mbox{$\overline{\textrm%
{Off}}$\hspace{.05em}\raisebox{.4ex}{$\underline{\textrm{line}}$}}\xspace}

\begin{document}

\preprint{Published in Phys. Rev. D as DOI: \href{http://dx.doi.org/10.1103/PhysRevD.93.122005}{10.1103/PhysRevD.93.122005}}

\title{Energy Estimation of Cosmic Rays with the Engineering Radio Array\\of the Pierre Auger Observatory}

\author{A.~Aab}
\affiliation{Universit\"{a}t Siegen, Fachbereich 7 Physik - 
Experimentelle Teilchenphysik, Siegen, 
Germany}
\author{P.~Abreu}
\affiliation{Laborat\'{o}rio de Instrumenta\c{c}\~{a}o e F\'{\i}sica Experimental 
de Part\'{\i}culas - LIP and  Instituto Superior T\'{e}cnico - IST, 
Universidade de Lisboa - UL, Lisboa, 
Portugal}
\author{M.~Aglietta}
\affiliation{Osservatorio Astrofisico di Torino  (INAF), 
Universit\`{a} di Torino and Sezione INFN, Torino, 
Italy}
\author{E.J.~Ahn}
\affiliation{Fermilab, Batavia, IL, 
USA}
\author{I.~Al Samarai}
\affiliation{Institut de Physique Nucl\'{e}aire d'Orsay (IPNO), 
Universit\'{e} Paris 11, CNRS-IN2P3, Orsay, 
France}
\author{I.F.M.~Albuquerque}
\affiliation{Universidade de S\~{a}o Paulo, Instituto de F\'{\i}sica, S\~{a}o 
Paulo, SP, 
Brazil}
\author{I.~Allekotte}
\affiliation{Centro At\'{o}mico Bariloche and Instituto Balseiro 
(CNEA-UNCuyo-CONICET), San Carlos de Bariloche, 
Argentina}
\author{P.~Allison}
\affiliation{Ohio State University, Columbus, OH, 
USA}
\author{A.~Almela}
\affiliation{Universidad Tecnol\'{o}gica Nacional - Facultad Regional 
Buenos Aires, Buenos Aires, 
Argentina}
\affiliation{Instituto de Tecnolog\'{\i}as en Detecci\'{o}n y 
Astropart\'{\i}culas (CNEA, CONICET, UNSAM), Buenos Aires, 
Argentina}
\author{J.~Alvarez Castillo}
\affiliation{Universidad Nacional Aut\'{o}noma de M\'{e}xico, M\'{e}xico, 
D.F., 
M\'{e}xico}
% ...
\collaboration{The Pierre Auger Collaboration}
\email{{\tt auger\_spokespersons@fnal.gov}}
\noaffiliation

\begin{abstract}
The Auger Engineering Radio Array (AERA) is part of the Pierre Auger Observatory and is used to detect the radio emission of cosmic-ray air showers. These observations are compared to the data of the surface detector stations of the Observatory, which provide well-calibrated information on the cosmic-ray energies and arrival directions. The response of the radio stations in the \hbox{\unit[30 to 80]{MHz}} regime has been thoroughly calibrated to enable the reconstruction of the incoming electric field. For the latter, the energy deposit per area is determined from the radio pulses at each observer position and is interpolated using a two-dimensional function that takes into account signal asymmetries due to interference between the geomagnetic and charge-excess emission components. The spatial integral over the signal distribution gives a direct measurement of the energy transferred from the primary cosmic ray into radio emission in the AERA frequency range. We measure \unit[15.8]{MeV} of radiation energy for a \unit[1]{EeV} air shower arriving perpendicularly to the geomagnetic field. This radiation energy -- corrected for geometrical effects -- is used as a cosmic-ray energy estimator. Performing an absolute energy calibration against the surface-detector information, we observe that this radio-energy estimator scales quadratically with the cosmic-ray energy as expected for coherent emission. We find an energy resolution of the radio reconstruction of 22\% for the data set and 17\% for a high-quality subset containing only events with at least five radio stations with signal. 
\end{abstract}

\pacs{96.50.sd, 96.50.sb, 95.85.Bh, 95.55.Vj}

\maketitle

\section{Introduction}
Cosmic rays in the ultra-high energy regime are detected through giant particle showers developing in Earth's atmosphere. Various detection systems are used to measure the calorimetric energy of the shower. Well established are techniques using telescopes that observe directly the shower development through fluorescence light emitted by molecules excited by the shower particles, and/or detectors positioned on the surface of Earth that measure the particles at one stage of the air-shower development (e.g., \cite{Auger2014, TA2008}). 

The observation of radio signals emitted by the shower particles using broadband megahertz (MHz) antenna stations has also been explored as a detection method to obtain complementary information on the air-shower development, and has become an active field of research in recent years \cite{LOPES2005, Codalema2006, HuegeRadioReview2014}. The properties of the primary cosmic rays have been studied in this way including their arrival direction, energy, and composition. Directional information can be obtained from the arrival times in several radio stations \cite{LOPES2005, Codalema2006, RAuger2012, Bezyazeekov201589}. To obtain information about the energy, calibrated detectors for cosmic-ray showers co-located with the radio stations are used \cite{GlaserARENA2012, LOPES_energyxmax_2014, Bezyazeekov201589}. Composition information has also been derived by relying on simulations of radio emission \cite{LOPES_energyxmax_2014, LofarXmaxMethod}.

One of the interesting characteristics of the radio-emission signal is the strong polarization of the electric field arriving at the antennas. Two components have been identified originating from different emission processes. The dominant one is perpendicular to Earth's magnetic field and is denoted as geomagnetic emission \cite{CodalemaGeoMag, LOPES2005, RAuger2012}. The second component is polarized radially with respect to the axis of the air shower and results from the negative charge excess in the shower front \cite{Hough1970, Prescott1971, Marin2011}. Its relative strength with respect to the geomagnetic emission is on average 14\% at the Auger site for an air shower arriving perpendicularly to the geomagnetic field \cite{AERAPolarization}. 

As a consequence of the superposition of the two emission mechanisms, the lateral distribution function (LDF) of the electric-field strength has been found to have a radial asymmetry \cite{WERNER2008, Ludwig2011438, ZHAires2012, LOFARLDF}. The two-dimensional shape of the LDF is best understood in a coordinate system with one axis perpendicular to the shower direction $\vec{v}$ and Earth's magnetic field $\vec{B}$ (along the Lorentz force $\sim \vec{v} \times \vec{B}$), and another along the perpendicular axis $\vec{v}\times (\vec{v}\times \vec{B})$. In this coordinate system the LDF exhibits a peanut-like shape.

So far, all radio experiments have used experiment-specific quantities to reconstruct the cosmic-ray energy, such as the radio signal strength at a characteristic lateral distance from the shower axis. 
While this method has long been known to provide a good precision \cite{Huege2008}, it has the disadvantage that the corresponding energy estimators cannot be directly compared across different experiments. The main reason for this is that the shape of the lateral signal distribution changes significantly with observation altitude. The optimal characteristic distance varies with observation height and even at the same characteristic distance the radio signal strengths are significantly different \cite{AERAEnergyPRL}.
Hence, a comparison between different experiments cannot be performed directly. 

In this contribution we introduce a general approach with a direct physical interpretation. At each observer position we calculate the energy deposit per area of the cosmic-ray radio pulse and by integrating the two-dimensional lateral distribution function over the area we obtain the total amount of energy that is transferred from the primary comic ray into radio emission during the air-shower development. This approach is independent of the shape of the signal distribution because energy, i.e., the integral over the signal distribution, is conserved.

In this analysis we present the relation between the cosmic-ray energy and the total energy emitted by the air shower as a radio pulse, for primaries of energy in the EeV (= \unit[10\textsuperscript{18}]{eV}) range. To obtain this relation we use radio stations of the Auger Engineering Radio Array located within the Pierre Auger Observatory in Argentina. The antennas are of the logarithmic periodic dipole antenna (LPDA) type and have been thoroughly studied and calibrated \cite{AntennaPaper}. We take advantage of the possibility to cross-calibrate these measurements with the well-understood data of the Observatory and with recent developments in understanding the radio-emission mechanisms, with their corresponding polarization patterns of the electric field and the particular lateral distribution of the total field strength \cite{CoREAS2013, LOFARLDF}.  

This paper is structured as follows. We begin with the experimental setup of the antenna array and the surface detector, then proceed to the data selection and event reconstruction procedure. After that we describe the calibration that uses a likelihood procedure, and we discuss experimental uncertainties. Finally, we present the energy measurement of the AERA radio detector, its resolution, the correlation of the radiation energy with the shower energy and we address the systematic uncertainty of the radio energy as an energy estimator.

A summary of the main results presented here and its implications on
the energy measurement of cosmic rays can be found in an accompanying
publication \cite{AERAEnergyPRL}.

\section{Detection systems}
The Pierre Auger Observatory is a hybrid detector for cosmic rays, based on two complementary detection systems. The surface detector (SD) array consists of 1660 water-Cherenkov detectors distributed over an area of \unit[3000]{km\textsuperscript{2}}. Its stations have a spacing of \unit[1.5]{km}, optimized to reach full efficiency for cosmic-ray energies above \unit[3]{EeV} \cite{Auger2014}. The fluorescence detector (FD) consists of 27 telescopes grouped at four locations around the area covered by the SD stations. With the FD, UV light is observed originating from the fluorescence emission of molecules excited by the cosmic-ray-induced air shower. 
The hybrid design of the Pierre Auger Observatory allows for an accurate energy calibration of the SD using the direct energy measurement of the FD. The amount of fluorescence light is proportional to the deposited energy and thus yields an accurate measurement of the energy of the primary particle.

The radio detector (RD) stations of AERA are located in an area of denser detector spacing of the SD array. This region, with SD station spacing of \unit[0.75]{km}, allows the detection of cosmic-ray energies down to about \unit[0.1]{EeV}.

The first deployment stage of AERA consists of 24 antenna stations with a spacing of \unit[144]{m}. Every station is equipped with two logarithmic-periodic dipole antennas \cite{AntennaPaper} integrated in one mechanical structure. The two antennas are oriented into the east-west and north-south directions relative to magnetic north. The corresponding analog and digital electronics are tuned to the frequency range of \unit[30 to 80]{MHz} \cite{KlausProceeding}. After filtering and amplification, the signal is digitized at \unit[180]{MSa/s} or \unit[200]{MSa/s} depending on the hardware type \cite{ARENA2014_Maller}. The stations are equipped with solar cells and a battery to ensure an autonomous power supply. Furthermore, all stations are connected via an optical fiber-network to the data-acquisition system (DAQ).

The system runs in two different modes, depending on the type of digitizing hardware. A self-trigger algorithm runs on the voltage trace itself, which identifies pulses based on characteristics described in \cite{Kelley2012}, and consequently creates a trigger. The triggers of multiple RD stations are checked for coincidences in a short time window of \unit[1]{$\upmu$s} -- compatible with the passage of an air shower -- at the DAQ level. A readout is requested once coincidences between at least three radio stations are found. Alternatively,  stations are triggered using an external trigger. Here, the DAQ receives a trigger from the Observatory's central data-acquisition system (CDAS) once an air-shower candidate has been registered with the SD or FD. 
This trigger initiates the readout of all the stations, which are equipped with a ring buffer. The buffer has a size of 4 GBytes and can store the traces of the two channels for about \unit[7.4]{s} which is sufficient to hold the data for the time needed to receive the trigger by the CDAS.

\section{Data selection and event reconstruction}
In this work we are using RD and SD data recorded between April 2011 and March 2013 when AERA was operating in its first commissioning phase. The data are stored as events, which refer to all relevant information that has been read out following a trigger. For this analysis, both self-triggered and externally triggered events are used. 

\subsection{Preselection of cosmic-ray candidates}
\label{preselection}
In the case of the self-triggered events, a preselection is performed offline by searching for coincidences with the surface detector events. A radio event has to agree in time and location with an SD event to be considered as cosmic-ray candidate. The radio-trigger time and the time when the air shower core hits the ground have to agree within $\pm$\unit[20]{$\upmu$s}. 
Such a conservative coincidence window also accounts for horizontal events, for which the time difference is expected to be larger.

For both trigger types, only events with a clear radio pulse 
in at least three stations are considered, to allow for a reconstruction of the incoming direction of the signal. 
For externally triggered events the requirement is a signal-to-noise ratio (SNR) greater than ten. Here the SNR is defined as the maximum of the Hilbert envelope-squared \footnote{The Hilbert envelope is the instantaneous amplitude.} divided by the noise variance. For self-triggered events the signal threshold is dynamically adjusted to the noise level to keep the trigger rate at a constant level of \unit[100]{Hz}.
We require that the reconstructed incoming directions from the radio and the surface detectors agree within 20\degree to be accepted as a cosmic-ray candidate. The 20\degree cut does not reflect the angular resolution of the SD nor that of the radio detector. This preselection cut retains the maximum number of cosmic-ray signals and significantly reduces the number of random (anthropogenic) noise pulses, which originate mainly from the horizon.

In addition, we apply quality cuts on the data of the surface detector \cite{AugerICRC2013Schulz}. The most important cuts are that the core position is closest to an active station and surrounded by a hexagon of active stations and that the zenith angle of the incoming direction be less than 55\degree. A total of 181 cosmic-ray candidates with energies above \unit[10\textsuperscript{17}]{eV} remain.

As an engineering array, AERA was subject to several changes in software and hardware which significantly limited the uptime. 
In future, we expect a larger rate of cosmic rays due to the stabilized operation of the detector.

\subsection{Reconstruction of radio data}
\label{chapter:reconstruction}
We use the software framework \Offline \cite{OfflineSoftware} of the Pierre Auger Collaboration to process the measured raw data. First, the air shower is reconstructed using the surface detector information \cite{AUGERICRC2011_Maris}. Second, the reconstruction using the radio detector data is performed \cite{OfflineRadio}. Narrowband noise sources are filtered out using a radio-frequency interference suppression in the time domain. Sine waves with the frequency of noise sources are fitted to the measured voltage trace and subtracted.

We correct for the influence of the analog signal chain using the absolute calibration of the AERA station and reconstruct a three-dimensional electric field by using the direction of the shower and applying the simulated antenna response \cite{AntennaPaper}.

\begin{figure}[btp]
\centering
\epsfig{file=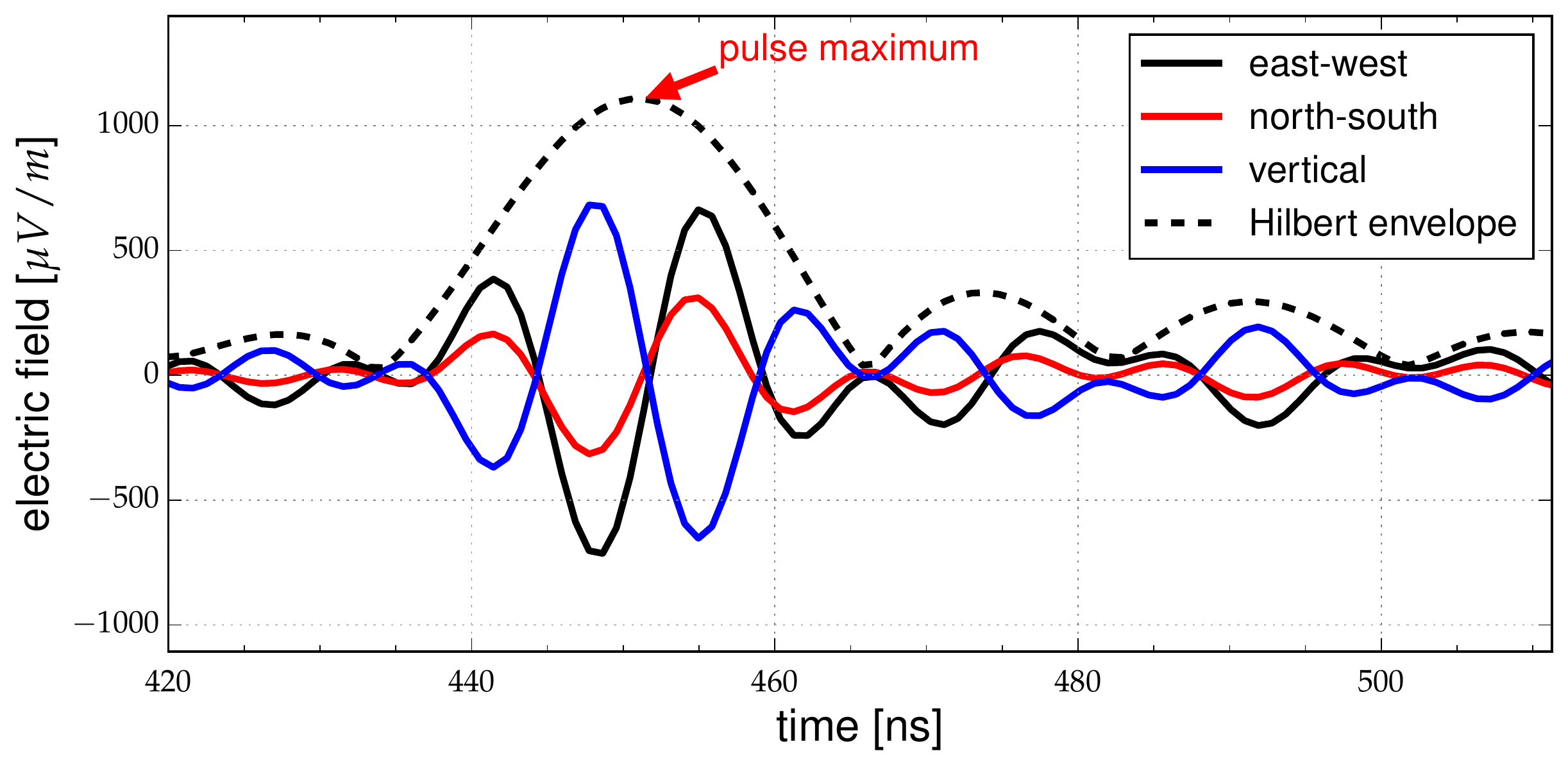, width=0.48\textwidth}
\caption{Reconstructed electric-field trace of one of the measured cosmic-ray radio events. An upsampling by a factor of five was applied. The shown Hilbert envelope (dashed line) is the square root of the quadratic sum of the Hilbert envelopes of the three polarization components.}
\label{fig:gEField}
\end{figure}

An example of a reconstructed electric-field trace $\vec{E}(t)$ is shown in Fig.~\ref{fig:gEField}. The energy fluence $f$, i.e., the energy deposit per unit area, of the incoming electromagnetic radio pulse at each radio station is determined by calculating the time integral over the absolute value of the Poynting vector.
This is achieved by squaring the magnitude of the electric-field trace and summing  over a time window of \unit[200]{ns} ($[t_1, t_2]$) around the pulse maximum which has been determined from the Hilbert envelope of the trace (cf. Fig.~\ref{fig:gEField}). The contribution of background noise (determined in the noise window $[t_3, t_4]$) is subtracted under the assumption that the main contribution is white noise. The energy fluence $f$ is given by
\begin{linenomath}\begin{equation}
f = \varepsilon_0 c \left( \Delta t \sum\limits_{t_1}^{t_2}|\vec{E}(t_i)|^2 - \Delta t \frac{t_2 - t_1}{t_4 - t_3} \sum\limits_{t_3}^{t_4}|\vec{E}(t_i)|^2  \right)\,,
\end{equation}\end{linenomath}
where $\varepsilon_0$ is the vacuum permittivity, $c$ is the speed of light in vacuum and $\Delta t$ is the size of one time bin. This quantity is used throughout the whole analysis and will be given in units of $\unit{eV/m^2}$.
To approximate the uncertainty the noise level as described above is used.
As the radio detector effects have been corrected for, the reconstructed energy fluence can be directly compared to air-shower simulations. 

We also calculate the direction of the electric-field vector, i.e., the polarization direction of the signal. In the full width half maximum (FWHM) interval around the pulse maximum of the Hilbert envelope we observe that the reconstructed electric-field vectors are aligned approximately along the same direction for every time bin. To accurately determine the mean direction of the electric-field vector, we average over all vectors in the FWHM interval of the Hilbert envelope (cf. Fig.~\ref{fig:gEField}).

\subsection{Selection of radio signals induced by cosmic rays} 
\label{cosmic_ray_selection}
Given the amount of pulsed background noise at the AERA site, the preselected events are likely to contain non cosmic-ray signals that mimic cosmic-ray pulses. There are two scenarios possible: 
Signals in one or more stations are not caused by the air shower or an event contains only noise pulses that by chance led to a reconstructed incoming direction similar to that of the SD. 

In order to reject background signals, we take advantage of the expected polarization of the radio signal. The polarization of the radio pulse is only used for this purpose and not considered for the energy estimation. In the frequency range of AERA (\unit[30 to 80]{MHz}) the dominant emission process is the geomagnetic emission \cite{CodalemaGeoMag, AERAPolarization}. Here, a linear polarization of the electric field is expected to be in the direction of the Lorentz force (given by $\vec{e}_{\mathrm{geo}}$) that acts on the charged particles while they traverse the magnetic field of Earth. The polarization is altered by an additional emission which is linearly polarized radially towards the shower axis (given by $\vec{e}_\mathrm{CE}$), and is referred to as the charge-excess emission process \cite{AERAPolarization, Askaryan1962, KahnLerche1966, Scholten2012}. 

The expected direction of the electric-field vector is therefore calculated from the geomagnetic and the charge-excess contributions
\begin{linenomath}\begin{equation}
 \vec{E}_\text{exp} \propto \sin\alpha \, \vec{e}_\text{geo} + a \, \vec{e}_\text{CE} \,,
 \label{eq:exp}
\end{equation}\end{linenomath} 
where $\alpha$ is the angle between the shower axis and magnetic field of Earth, and $a$ is the average relative charge-excess strength that has been measured to be $0.14 \pm 0.02$ at AERA \cite{AERAPolarization}. In this approach, the direction of the geomagnetic contribution depends only on the incoming direction of the air shower whereas the charge-excess contribution depends in addition on the position of the radio station relative to the shower axis. 

\begin{figure}[btp]
  \centering
    \epsfig{file=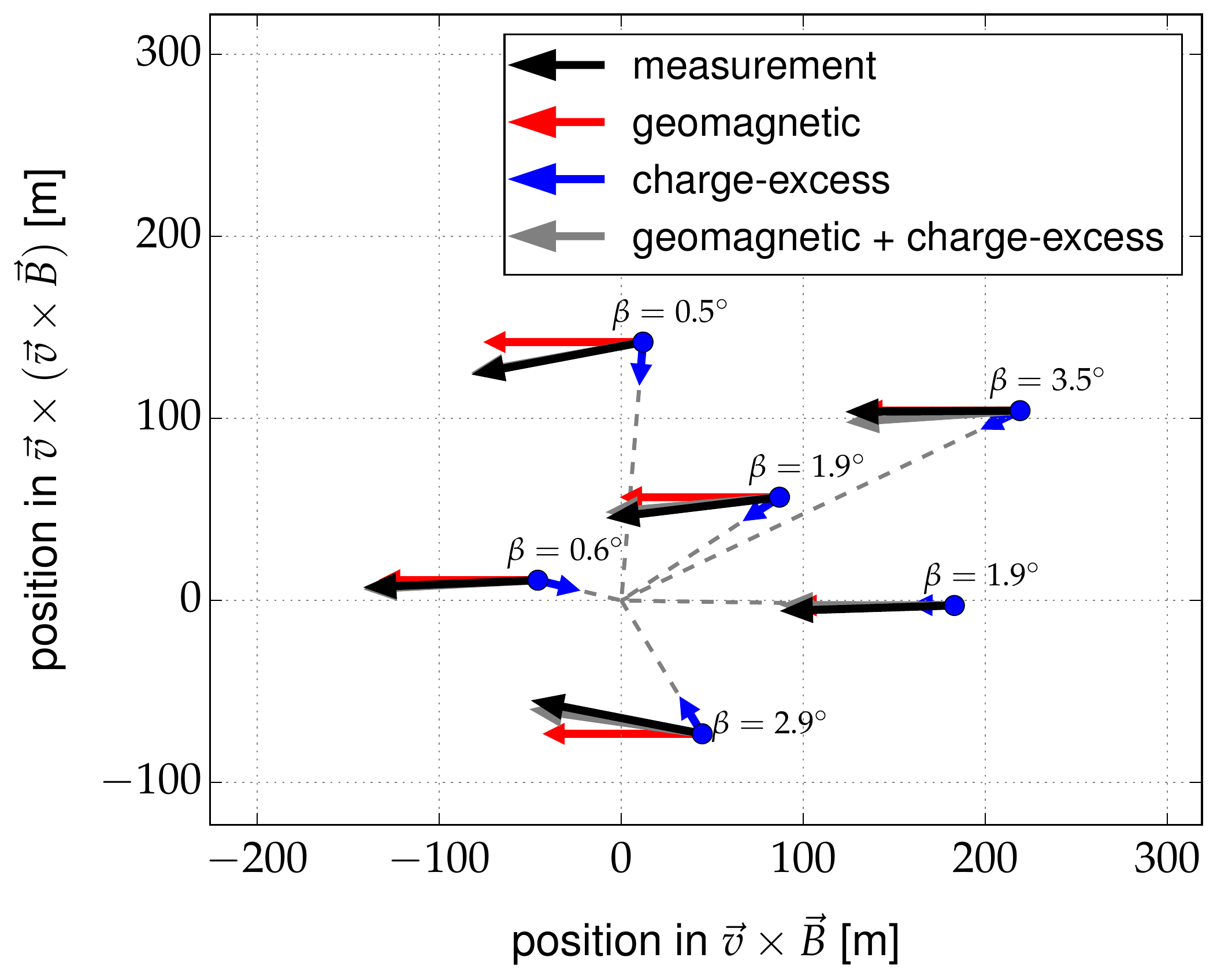, width=0.48\textwidth}
  \caption{Polarization map of a single event. The axis coordinates are in the shower plane where the x-axis corresponds to the direction of the Lorentz force ($\vec{v} \times \vec{B}$) and the y-axis perpendicular to that and to the shower axis ($\vec{v} \times (\vec{v} \times \vec{B})$). The SD shower core is at the coordinate origin. The measured polarizations are shown as the black arrows. The gray arrows are the model expectations, and the red and blue arrows are the geomagnetic and the charge-excess components, respectively. The definition of $\beta$ is described in the text. The air-shower properties of this event are: Energy of \unit[0.9]{EeV} and arriving from a zenith angle of 36\degree and from 27\degree south of west. For the emission model of Eq.~\eqref{eq:exp}, the optimal value of the relative charge-excess strength is $a = 0.18$.}
  \label{fig:polarizationmap}
\end{figure}

In Fig.~\ref{fig:polarizationmap}, all stations with signal of a cosmic-ray candidate are shown, and the measured polarization is compared with the expectations of the two radio-emission mechanisms. The overall agreement between measured and expected field polarizations is quantified using the angular difference

\begin{linenomath}\begin{equation}
\beta_i = \angle(\vec{E}_{\text{meas},i}, \vec{E}_{\text{exp},i})
\end{equation}\end{linenomath} 
at each station $i$. For each event, the average deviation $\bar{\beta}$ of the individual deviations $\beta_i$ of the stations with signal is calculated and will be used as criterion for a quality cut. Relevant uncertainties are taken into account as follows: 
\begin{itemize}
\item The relative strength $a$ of the charge excess can vary from event to event due to shower-to-shower fluctuations, and additional dependencies on the geometry of the air shower \cite{LofarPolarization2014}. Therefore, for each possible values of $a$ between $0$ and $0.5$ the average deviation $\bar{\beta}$ is calculated and only the smallest value of $\bar{\beta}$ is considered. 
\item The uncertainty of the SD shower core position is taken into account by variation of the core within its estimated uncertainties. In our data set the uncertainty varies between \unit[10]{m} and \unit[80]{m} depending on the energy and zenith angle.  For each trial of the core position $\bar{\beta}$ is calculated. Again, only the smallest value of $\bar{\beta}$ is considered.
\item Interference of the cosmic-ray radio signal with noise pulses can alter the polarization. Simulation studies showed that for a single radio station the uncertainty in $\beta$ due to noise is below 8\degree at detection threshold, and decreases to 1\degree at high signal-to-noise ratios. To obtain the average value of $\beta$ for all radio stations in the event
we compute a weighted mean with weights $w_i = 1/\sigma^2_{\beta_i}$ with $\sigma_{\beta_i}$ being the expected uncertainty from the simulation. 
\end{itemize}

We impose a limit on the average deviation $\bar{\beta}$ of the polarization direction. This maximum deviation is fixed at a value of 3\degree. This value is slightly above the combination of the following effects.

The incoming direction of an air shower reconstructed with the surface detector has an uncertainty between 1.3\degree and 0.7\degree depending on the cosmic-ray energy and the zenith angle \cite{AUGERICRC2011_Maris}. Hence, the expected direction of the electric-field vector will have the same uncertainty. 
All antennas are aligned to the magnetic north (or perpendicularly to the magnetic north in case of the other polarization direction) with a precision of better than 1\degree \cite{PHDFliescher}. 
All antennas are uniformly constructed and the two antennas of a radio station are identical. Asymmetries in the ground conditions have only negligible influence as the LPDA antenna is mostly insensitive towards the ground. A measurement at AERA has shown that the responses of all antennas differ by less than 0.3\% \cite{StephanAsiaPacificConference2010}.  

A difference in the amplification of the signal chain of the north-south and east-west polarized antenna will influence the polarization measurement. From an individual measurement of the signal chain of all antennas the uncertainty is estimated to be 2.5\% which results in a polarization uncertainty below 0.7\degree. 

In addition, we neglect the dependence of the relative strength $a$ of the charge excess on the distance between observer position and shower axis \cite{Vries2010a, LofarPolarization2014}. 
For a single station this effect is relevant. However, in our approach we only use the average deviation of all stations with signal also taking into account the uncertainty in the core position. Therefore the distance dependence will mostly average out.
We estimate that the remaining additional scatter is 1.5\degree.

We account for individual radio stations being contaminated with substantial noise signals by iterating through all configurations with only one and then more stations removed, down to the minimum of three stations. 
An event where the weighted average deviation $\bar{\beta}$ is greater than 3\degree for all station combinations is rejected. 
If $\bar{\beta}$ is less than 3\degree for any station combination and the fraction of selected stations is larger than 50\% of the total number of stations with signal, the event candidate is considered a cosmic-ray event and only the stations from this particular combination are used. After this cut 136 events remain. The number of excluded single stations and complete events is compatible with the measured rate of noise pulses. 

Most of the events recorded during thunderstorm conditions appear to be rejected by this selection procedure as the strong atmospheric electric fields of a thunderstorm influence the radio emission and alter the polarization of the radio signals \cite{ICRC2009_Thunderstorm, LofarThunderstorm2015}. For two thirds of the events, a measurement of the atmospheric electric field is available. These events are checked for thunderstorm conditions using an algorithm described in \cite{Nehls2008}. Based on this check, two additional events were rejected. All cuts are summarized in Table \ref{tab:cuts}.

\begin{table}[htb]
\caption{\label{tab:cuts}Overview of selection cuts and the number of events surviving these cuts. Preselection means: E\textsubscript{CR} $\geq$ $\unit[0.1]{EeV}$, standard SD quality cuts, $\geq$ 3 radio stations with signal, SD and RD reconstructed incoming directions agree within 20\degree. See text for details.}
 \centering
 \begin{tabular}{p{0.32\textwidth} p{0.15\textwidth}}
 \hline \hline
  cut & number of events after cut \\ \hline
  preselection (Sec. \ref{preselection}) & 181  \\
  polarization cut ($\bar{\beta} <$ 3\degree, Sec. \ref{cosmic_ray_selection}) & 136  \\
  no thunderstorm conditions (Sec. \ref{cosmic_ray_selection}) & 134  \\
  LDF fit converged ($\sigma <$ \unit[300]{m}, Sec. \ref{LDF})  & 126  \\
  $\geq$ 5 stations with signal \\ (only high-quality data set, Sec. \ref{energy_calibration}) & 47  \\
  \hline \hline
 \end{tabular}
 \end{table}

\subsection{Uncertainties on the energy fluence in a single radio station}
In addition to the uncertainties on the amplification of the signal chain of 2.5\% discussed above, no further uncertainties are expected that would result in a different response of stations within one event. To first order, the frequency content and the incoming direction of the radio pulse are similar at all observer positions. Therefore, an uncertainty of the antenna-response pattern has a negligible influence as it is evaluated for the same direction at all stations. Possible different ground conditions at different station positions that result in a different reflectivity of the soil are negligible due to the insensitivity of the antenna towards the ground. 
The 2.5\% amplification uncertainty results in 5\% uncertainty on the energy fluence $f$, as $f$ scales quadratically with the electric-field amplitude. This uncertainty is added in quadrature to the signal uncertainty resulting from noise.

\section{Energy Estimator}
\label{LDF}

\begin{figure*}[tbp]    
    \includegraphics[height=0.34\textwidth]{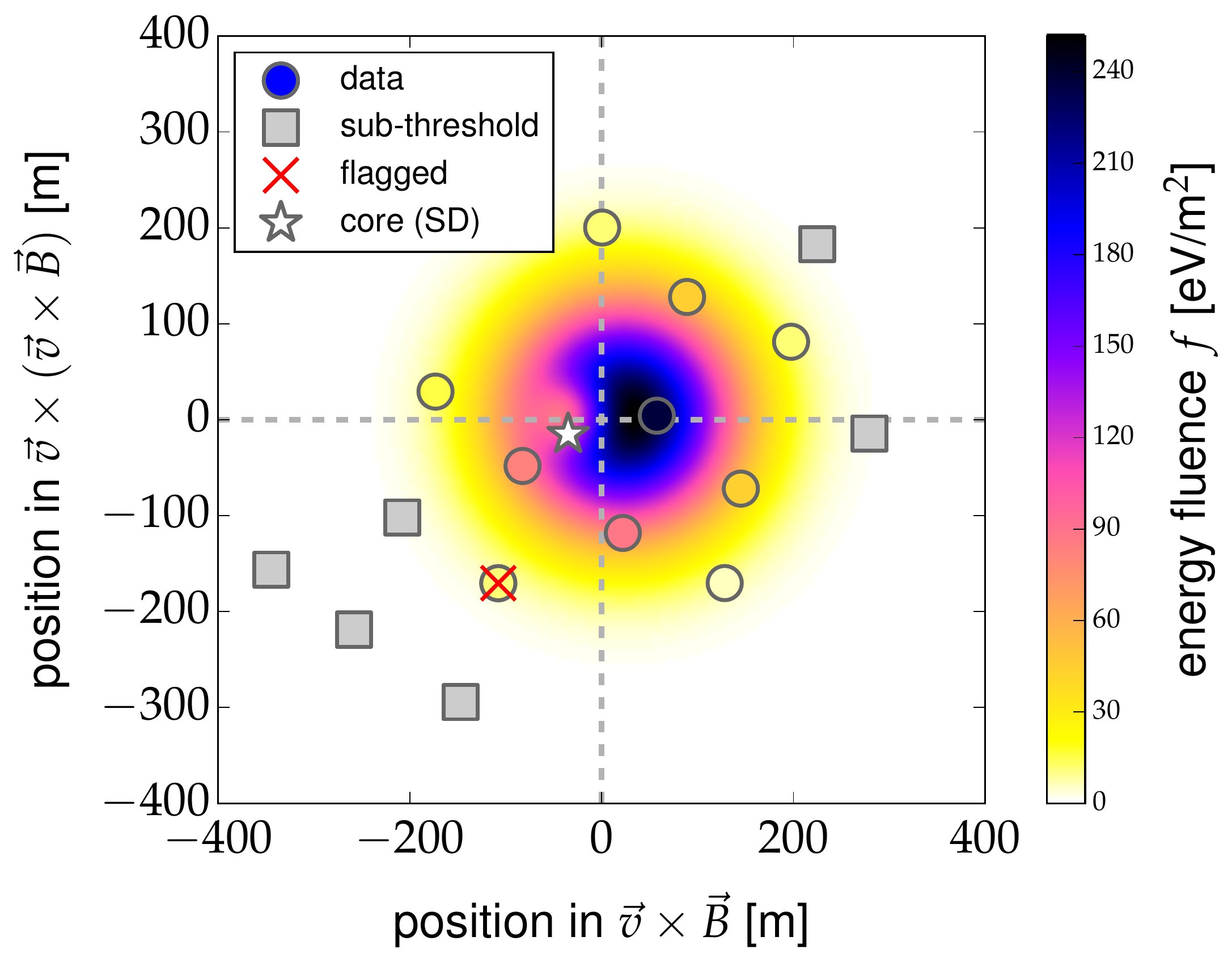} 
    \hspace{0.05\textwidth}
        \includegraphics[height=0.34\textwidth]{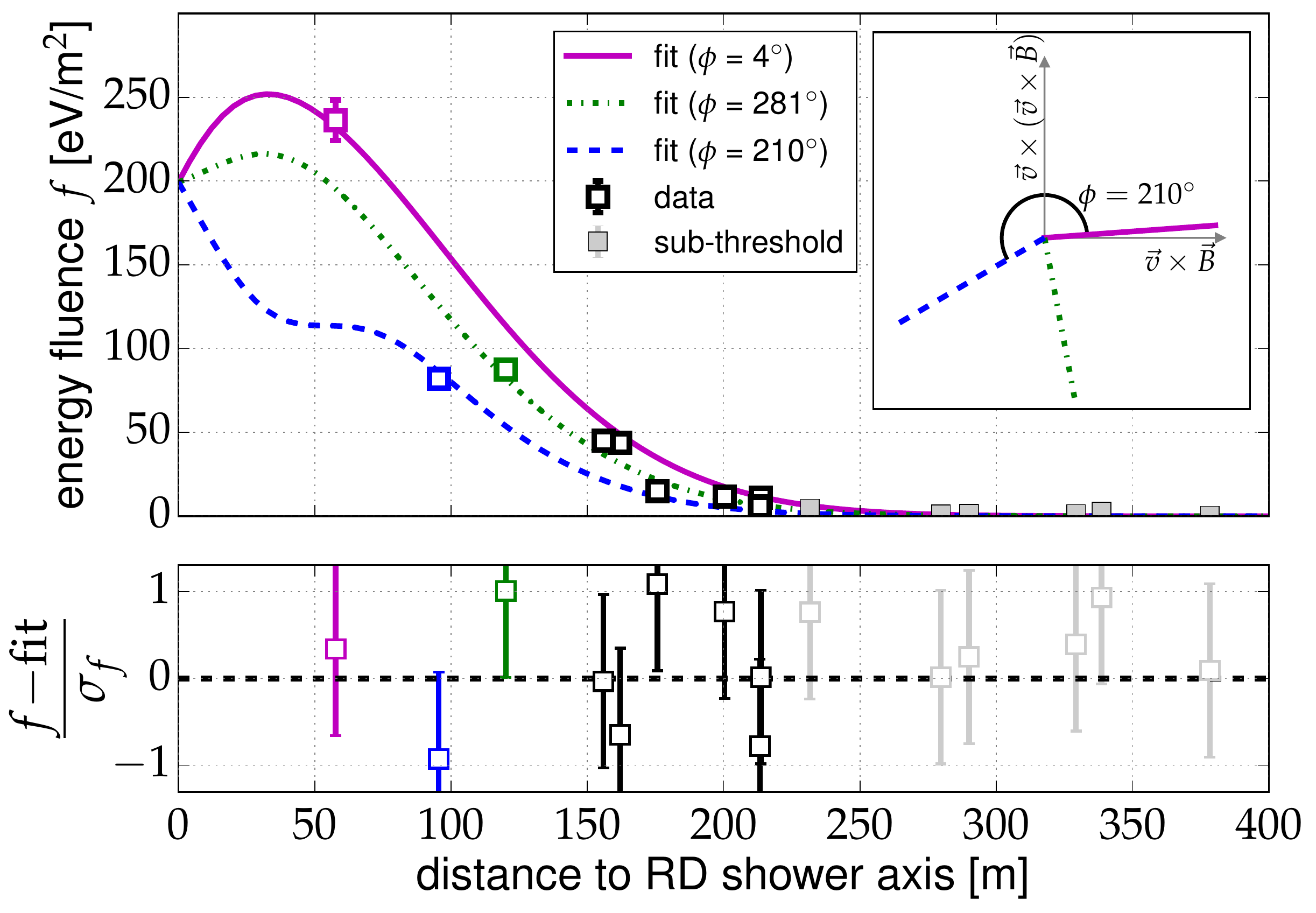} 
\caption{Lateral signal distribution of a single cosmic-ray event. The air-shower properties of this event are: Energy of \unit[0.75]{EeV} and arriving at a zenith angle of 37\degree and from 44\degree west of south. Left: The energy fluence in the shower plane. The measurements are indicated as circles where the color shows the energy fluence. Gray squares are stations with signal below threshold and the red cross marks a station that is rejected due to a mismatch in the signal polarization. The background map shows the LDF parametrization. The coordinate origin is the reconstructed core position of the radio LDF fit. Note the lack of color contrast between the infill color of the data points and the background. This is indicative of the agreement between the data and the model. Right: Representation of the same data as a function of distance from the shower axis. The colored and black squares are the measured energy fluences and gray squares are the stations with signal below threshold. For the three data points with the highest energy fluence, the one-dimensional projection of the two-dimensional LDF onto lines connecting the radio-core position with the corresponding radio detector positions is illustrated with colored lines. This demonstrates the azimuthal asymmetry and complexity of the two-dimensional lateral distribution function. The inset figure shows the azimuthal direction of the three LDF projections. The distribution of the residuals (data versus fit) is shown as well.}
\label{fig:LDF}
\end{figure*}

To obtain an absolute energy estimator from the signals at the different distances to the shower axis (energy fluence $f$ in units of \unit{eV/m$^2$}) a LDF is used which takes into account the signal asymmetries due to constructive and destructive interference between the geomagnetic and charge-excess components, as well as Cherenkov time-compression effects \cite{LOFARLDF}. This LDF describes the main features seen in simulated and measured cosmic-ray radio events. The LDF function is parametrized as
\begin{linenomath}\begin{equation}
\begin{split}
f(\vec{r}) =&\,A \Biggr[ \, \exp\left(\frac{-(\vec{r} + C_1 \, \vec{e}_{\vec{v}\times\vec{B}} - \vec{r}_\text{core})^2}{\sigma^2}\right) \\
 &- C_0 \, \exp\left(\frac{-(\vec{r} + C_2 \, \vec{e}_{\vec{v}\times\vec{B}} - \vec{r}_\text{core})^2}{(C_3 e^{C_4 \, \sigma})^2}\right) \Biggr] \,.
\end{split}
\label{eq:LDF}
\end{equation}\end{linenomath}
All coordinates are in the shower plane. $\vec{r}$ denotes the station position. The four fit parameters are the amplitude $A$, the slope parameter $\sigma$ and the particle core position $\vec{r}_\text{core}$.
In case of low station multiplicity, the particle core position is taken from the SD reconstruction, which enables us to also use events with only three or four stations with signal. $C_0$ - $C_4$ are constants that are estimated from CoREAS Monte Carlo simulations \cite{CoREAS2013} and can be found in Appendix \ref{LDF_parametersr}. $C_0 - C_2$ are zenith-angle dependent.
The LDF is fitted to the data using a chi-square minimization. An example of one air shower within our data set is shown in Fig.~\ref{fig:LDF}.

Some events do not contain sufficient information to fit the LDF, such as when only three stations with signal are present that have roughly the same signal strength. This results in an unphysically broad LDF. To reject these events we impose the quality cut $\sigma <$ \unit[300]{m} (Table \ref{tab:cuts}). An analysis of air-shower simulations for the AERA geometry showed that the $\sigma$ parameter of the LDF is never larger than \unit[300]{m}. 

In the following, only the 126 events that pass the quality cuts are considered and will be referred to as the full data set. To derive the accuracy of the energy estimation method, the data set will be further divided in a high-quality data set containing only events with at least five stations with signal, i.e., events where the core position can be reconstructed in the radio LDF fit. 

\begin{figure}[btp] 
	\centering   
\includegraphics[width=0.48\textwidth]{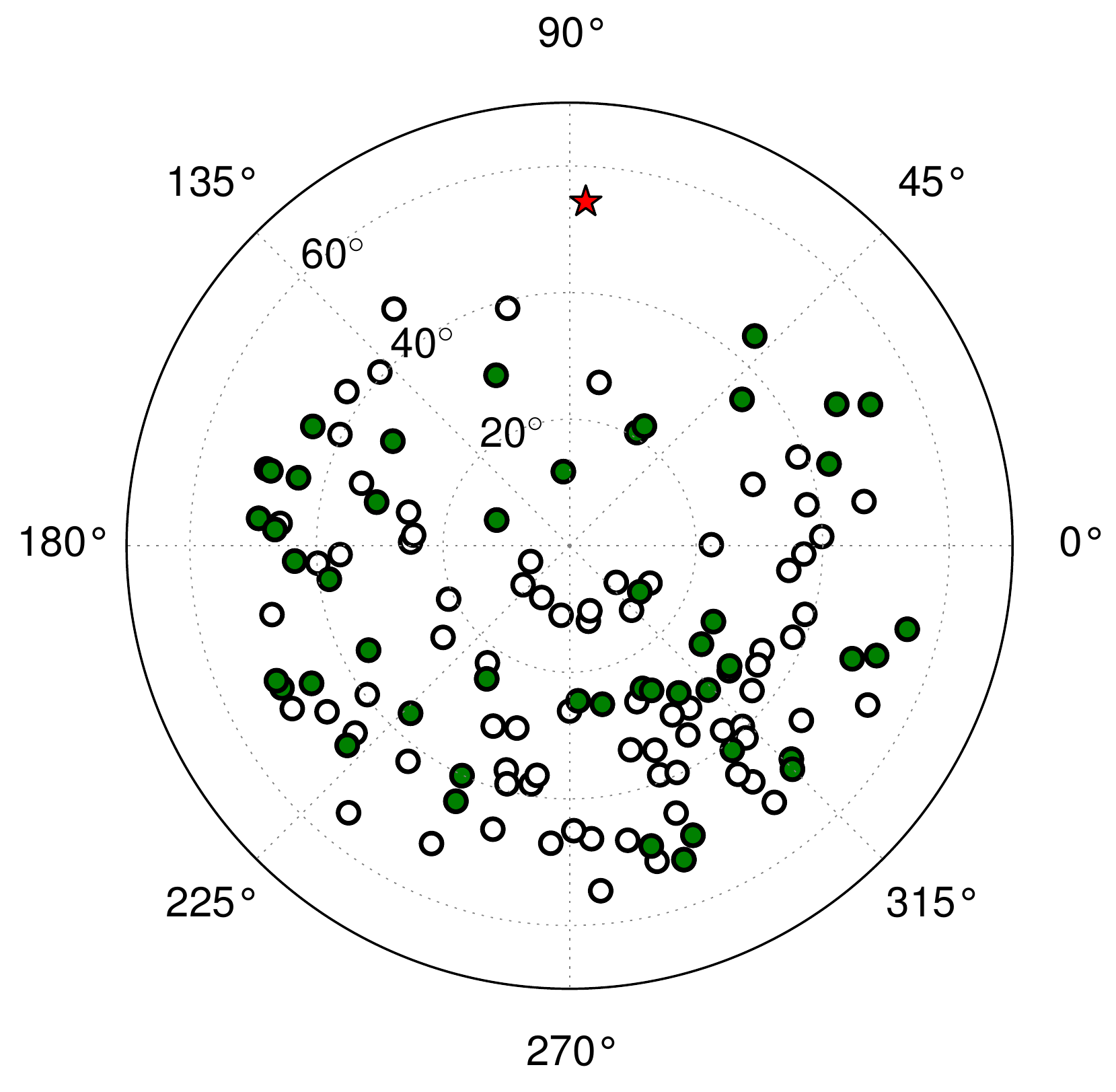} 
\caption{Skymap of the 126 selected events. Green filled circles denote air showers with at least five stations with signal and open circles denote air showers with less than five stations with signal. The red star denotes the direction of the magnetic-field axis at AERA. All measured events are at least 20\degree away from the magnetic-field axis. Therefore, the geomagnetic emission gives the dominant contribution to the radiation energy for all events. 
}
\label{fig:skymap}
\end{figure}

\subsection{Definition of the energy estimator}
The spatial integral of the lateral distribution function gives the amount of energy that is transferred from the primary cosmic ray into radio emission in the AERA frequency band during the air-shower development, and will be given in units of \unit{eV}. We define the energy estimator $S_\text{radio}$ as this radiation energy divided by $\sin^2\upalpha$ to account for different emission strengths at different angles between shower axis and magnetic field, see Eq.~\eqref{eq:exp},
\begin{linenomath}\begin{equation}
\begin{split}
S_\text{radio} &= \frac{1}{\sin^2\alpha}\int\limits_{\mathbb{R}^2} f(\vec{r}) \, \mathrm{d}^2\vec{r} \\ &= \frac{A \pi}{\sin^2\alpha} \,\left(\sigma^2 - C_0 C_3^2 \, e^{2 C_4 \sigma}\right) \,,
\end{split}
\label{eq:sradio}
\end{equation} \end{linenomath}
where $\mathbb{R}^2$ denotes the shower plane. The positive $\sigma^2$ term dominates by far over the negative second term resulting in a positive value of $S_\mathrm{radio}$. 
The $\sin^2\upalpha$ correction only holds if the geomagnetic emission is the dominant contribution which is the case for $\upalpha >$ 10\degree at AERA. Due to the reduced emission strength the number of detections for arrival directions within 10\degree of the geomagnetic field axis is suppressed. The angular distribution of the events is shown in Fig. \ref{fig:skymap}. 

\subsection{Event-by-event uncertainties of the energy estimator}
\label{eventbyeventuncertainties}
\begin{table*}[tbp]
\caption{\label{tab:uncertainties}Overview of uncertainties of the electric-field amplitude $\sigma_{|\vec{E}|}$ and the energy estimator $S_\text{radio}$. ``$\oplus$'' denotes a quadratic sum. The average fit uncertainty of $S_\mathrm{radio}$ is 46\%, and 24\% for the high-quality subset of events with at least five stations with signal. } 
 \begin{tabular}{lcc} 
 \hline \hline 
  \textbf{source of uncertainty} & $\boldsymbol{\sigma_{|\vec{E}|}}$ &$\boldsymbol{\sigma_{S_\text{radio}}}$\\ \hline
  \hspace*{0.55cm}\textbf{event-by-event} & \\
  \hspace*{0.55cm}\hspace*{0.55cm} temperature dependence  & 4\% & 8\% \\
  \hspace*{0.55cm}\hspace*{0.55cm} angular dependence of antenna response pattern & 5\% & 10\% \\
  \hspace*{0.55cm}\hspace*{0.55cm} reconstructed direction & negligible & negligible \\ 
    \hspace*{0.55cm}\hspace*{0.55cm} LDF fit uncertainty & - &error propagation of fit parameters \\ 
\hspace*{0.55cm}\textbf{total event-by-event uncertainty}  & \textbf{6.4\%} &\textbf{12.8\%} $\boldsymbol\oplus$ \textbf{fit uncertainty}  \\ \noalign{\smallskip}
  \multicolumn{3}{l}{\textbf{\hspace*{0.55cm}absolute scale}} \\
  \hspace*{0.55cm}\hspace*{0.55cm} absolute scale of antenna response pattern & 12.5\%  &  25\% \\
  \hspace*{0.55cm}\hspace*{0.55cm} analog signal chain & 6\% & 12\% \\ 
  \hspace*{0.55cm}\hspace*{0.55cm} LDF model & $<$2.5\% & $<$5\% \\ 
\hspace*{0.55cm}\textbf{total absolute scale uncertainty}  & \textbf{14\%} &  \textbf{28\%} \\ \hline \hline
 \end{tabular}
\end{table*}

The following uncertainties are relevant for the energy estimator due to event-by-event fluctuations and summarized in Table \ref{tab:uncertainties}: 
\begin{itemize}
\item The gains of the low-noise amplifiers and filter amplifiers exhibit a temperature dependence. The effect has been measured and amounts to $\unit[-42]{mdB/K}$. Each air shower is measured under specific environmental conditions. In particular this implies that we have a random distribution of ambient temperatures which exhibit a Gaussian distribution with a standard deviation of 8.3\degree C. This corresponds to a fluctuation of the gain of 4\%.
\item An uncertainty of the simulated antenna response that depends on the incoming direction of the radio signal will lead to an event-by-event uncertainty as each event has a different incoming direction. The effect is determined to be 5\% by comparison of the simulated antenna response with a measurement at AERA \cite{AntennaPaper}.  
\item The reconstructed direction of the air shower obtained with the SD has an uncertainty of less than 1.3\degree. This has negligible influence on the antenna response pattern, since it can be considered uniform over such a small change of angle.  
\end{itemize} 
As the different uncertainties are independent, the total uncertainty of the electric-field amplitude is $\sqrt{4\%^2 + 5\%^2} \approx 6.4\%$ and therefore 12.8\% on $S_\text{radio}$. The uncertainty of $\upalpha$ can be neglected. 
The fit uncertainties of $A$ and $\sigma$ including their correlation are propagated into $S_\text{radio}$ using Gaussian error propagation. In the case of events with less than five stations with signal, the core position of the surface detector reconstruction is used and its uncertainty is propagated into the fit uncertainty of $S_\text{radio}$. This fit uncertainty is added in quadrature to the statistical uncertainty of 12.8\% of the energy estimator. The average fit uncertainty of $S_\mathrm{radio}$ is 46\%. For events with at least five stations with signal the average uncertainty reduces to 24\%. 

\subsection{Absolute scale uncertainties of the energy estimator}
\label{systematics_sradio}
The dominant systematic uncertainties of the reconstructed electric-field amplitudes are the calibration of the analog signal chain and the antenna response pattern. The analog signal chain consists of the low-noise amplifier, the filter amplifier and all cables between the antenna and the analog-to-digital converter. The analog signal chain has been measured for each channel of each radio station separately in the field and differences are corrected for. The systematic uncertainty of the analog chain amounts to 6\%. 

The simulated antenna response pattern has been confirmed by measurements at an overall level of 4\%. The systematic uncertainty of the measurement is 12.5\% in the vector effective length \cite{PHDWeidenhaupt}. Conservatively, the systematic uncertainty of the antenna-response pattern is therefore estimated as 12.5\%. 

Systematic uncertainties introduced by the usage of the two-dimensional signal distribution function of Eq.~\eqref{eq:LDF} are negligible. Detailed comparisons of the shape of the radio signal distribution measured with LOFAR with the predictions from CoREAS show no indication of any systematic discrepancy \cite{LOFARNature2016}. We determined the influence of the 2D-LDF model on the radiation energy in a representative CoREAS Monte Carlo data set for the AERA detector and found a systematic effect of less than 5\%.

Combining all uncertainties in quadrature, the systematic uncertainty of the electric-field amplitude is 14\%. The radio-energy fluence and the energy estimator scale with the amplitude squared. Therefore, the systematic uncertainty of the absolute scale of the radiation energy is 28\%. We note that, as the cosmic-ray energy is proportional to the square root of the radiation energy (see next section), the systematic uncertainty of a radio cosmic-ray energy scale would remain at 14\%.

\section{Energy calibration}
\label{energy_calibration}

\begin{figure}[tbh]
\subfigure{
  \includegraphics[width=0.48\textwidth]{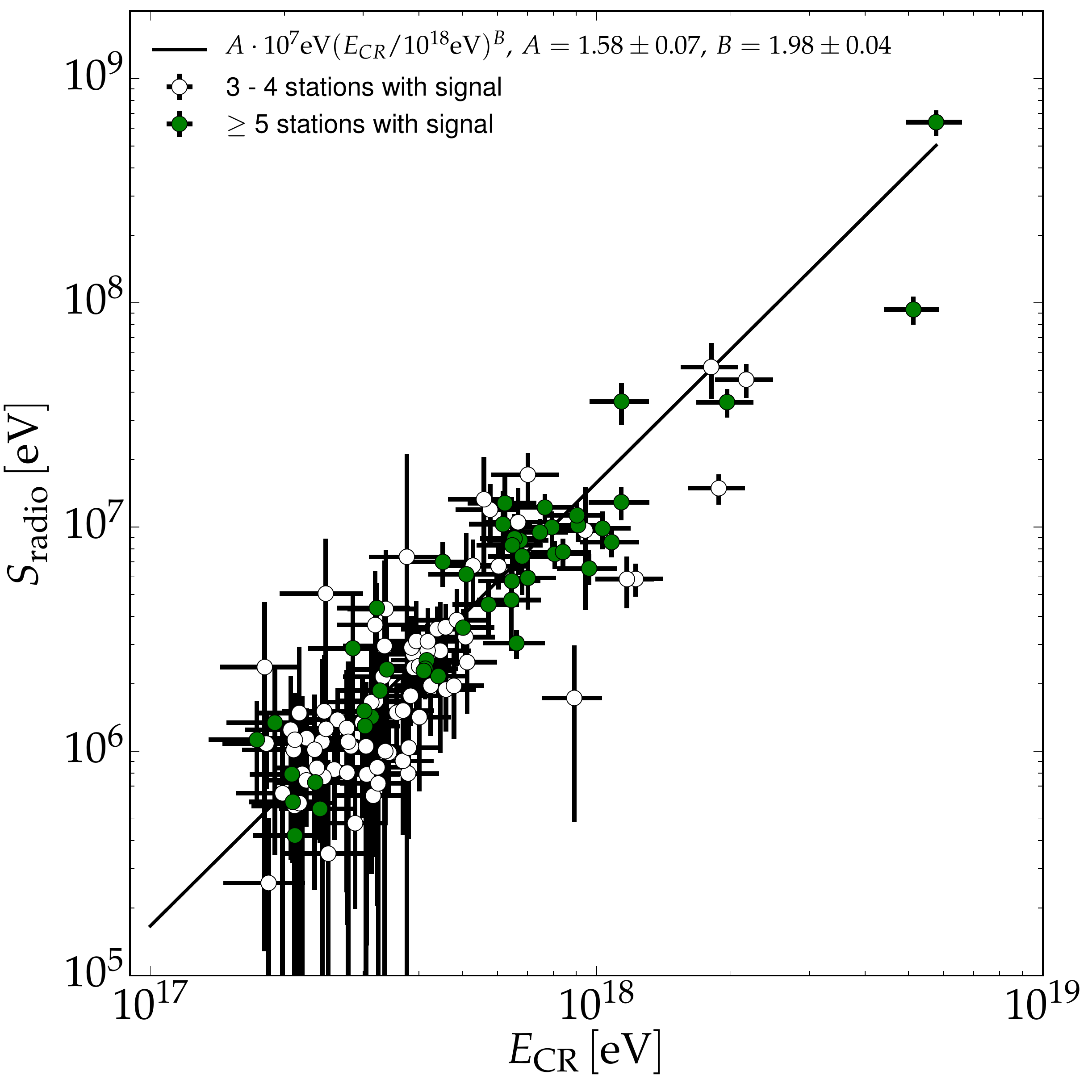}
}
\subfigure{
  \includegraphics[width=0.23\textwidth]{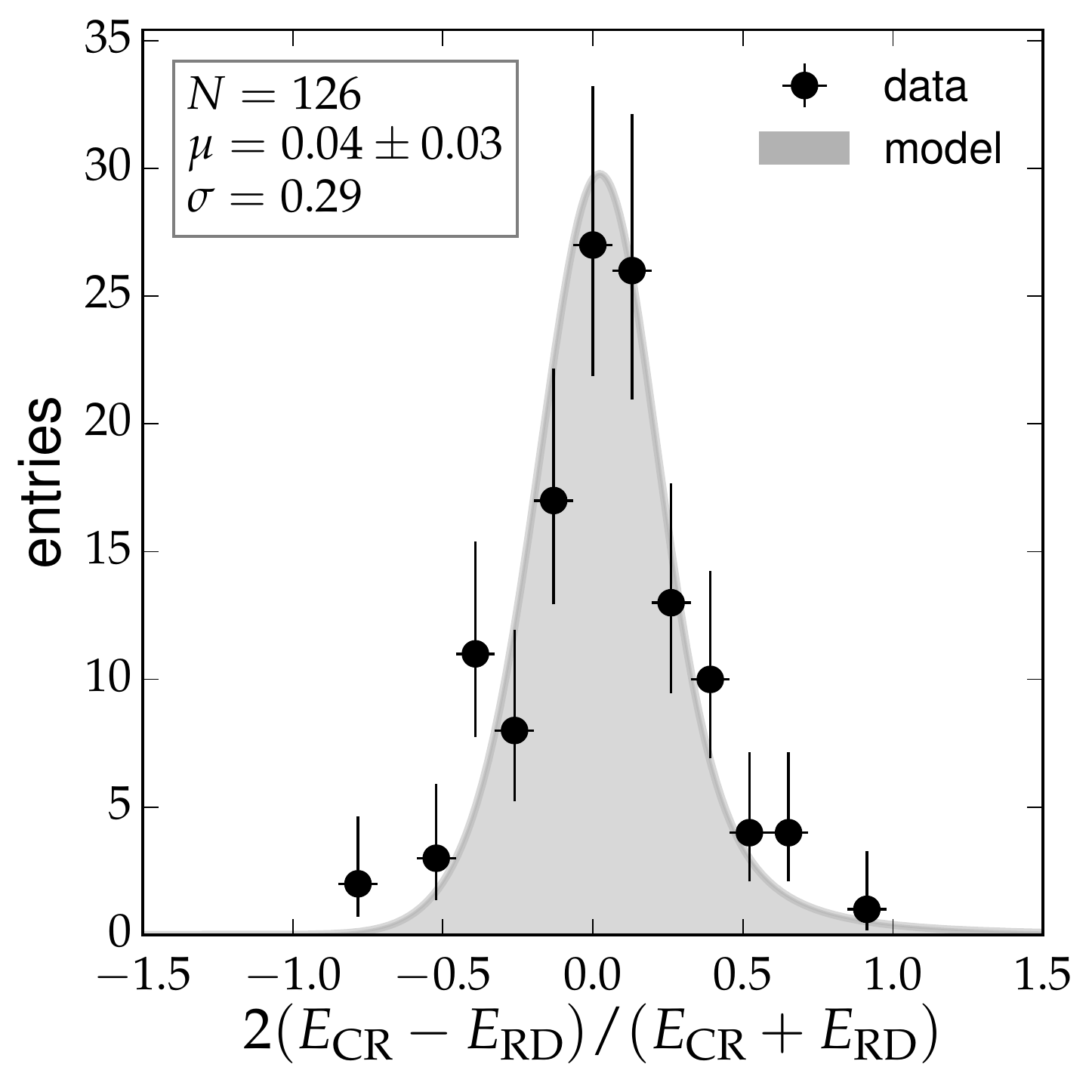}
  \includegraphics[width=0.23\textwidth]{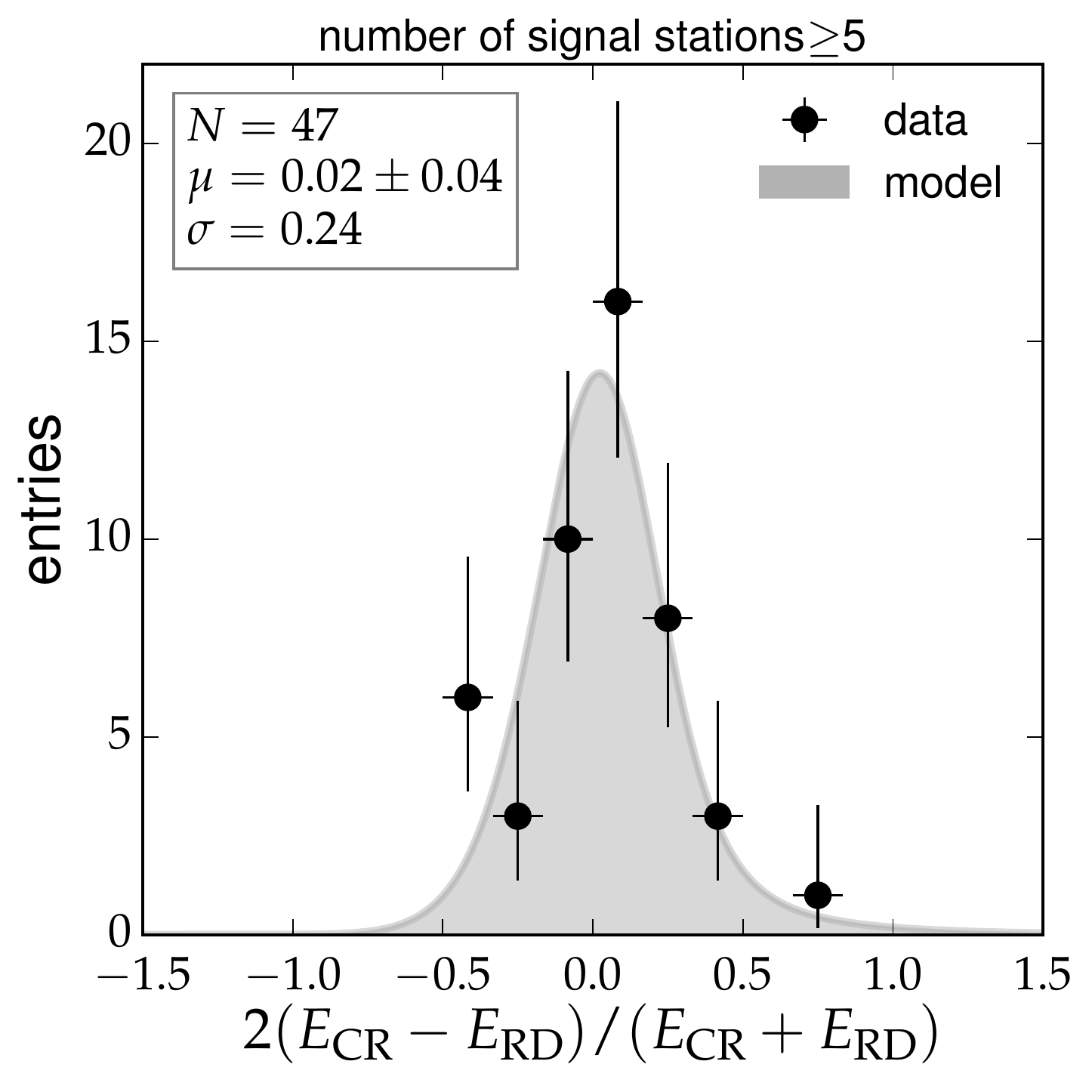}
}
\caption{(top) The radio-energy estimator $S_\text{radio}$ as a function of the cosmic-ray energy $E_\mathrm{CR}$ measured with the surface detector. A power law is fitted to the data using a likelihood approach which takes all uncertainties and detection efficiencies into account. Green filled circles denote air showers where the core position has been determined in the radio LDF fit, i.e., all air showers with at least five stations with signal. Open circles denote events with less than five stations with signal and use the SD core position. (bottom) Relative energy resolution: The energy of the radio detector is obtained using the fit in the left-hand figure. The left histogram contains all air showers, and the right histogram contains the air showers with at least five stations with signal (green filled circles). The expected distribution is shown as a gray shaded area which is computed from the fitted probability model that describes the fluctuations.}
\label{fig:EnergyCalibration}
\end{figure}

The radio-energy estimator $S_\text{radio}$ is shown as a function of the cosmic-ray energy $E_\mathrm{CR}$ measured with the surface detector in Fig.~\ref{fig:EnergyCalibration} top. A clear correlation is observed.
For the calibration function we follow the same method as used for the calibration of surface detector events with fluorescence detector events of the Pierre Auger Observatory \cite{AugerEnergyCalibration2013, N19_2015, Dembinski2015}. The calibration function 
\begin{linenomath}\begin{equation}
\label{eq:calibration_sradio}
S_\text{radio} = A \times \unit[10^7]{eV} \,(E_\text{CR}/\unit[10^{18}]{eV})^B
\end{equation}\end{linenomath}
is obtained by maximizing a likelihood function that takes into account all measurement uncertainties, detector efficiencies and the steeply falling energy spectrum (the functional form of the likelihood function can be found in appendix \ref{Likelihood}). The result of the calibration fit is $A = 1.58 \pm 0.07$ and $B = 1.98 \pm 0.04$. The correlation between $A$ and $B$ is 35\%.
The resulting slope is quite compatible with an exponent of $B=2$ implying that the energy deposited in radio emission increases quadratically with the cosmic-ray energy. If $B$ is fixed to $2$ the fit result is  $A = 1.59 \pm 0.06$. We can infer from Eq.~\eqref{eq:calibration_sradio} that, for a \unit[1]{EeV} air shower perpendicular to the magnetic field axis, \unit[15.8]{MeV} is deposited on average in radio emission in the frequency range of \hbox{\unit[30 to 80]{MHz}}.

The lower left panel of Fig.~\ref{fig:EnergyCalibration} shows the scatter around the calibration curve for all air showers in our data set. This amounts to 29\%. 
We also tested a high-quality data set containing only air showers with at least five stations with signal, where a determination of the core position in the radio LDF fit is possible. These air showers are marked by green filled circles in Fig.~\ref{fig:EnergyCalibration}. The fit of the calibration curve gives a compatible result ($A = 1.60 \pm 0.08,\, B=1.99 \pm 0.05$) and the scatter around the calibration curve reduces to 24\% (lower right panel of Fig.~\ref{fig:EnergyCalibration}). 

To obtain a goodness-of-fit estimator, the measured distribution is compared to the expected distribution which is computed from the likelihood function, i.e., from the probability model that describes the fluctuations. The comparison yields a reduced chi-square value of $\chi^2/\mathrm{ndf} = 13.8/12$ for the full data set and $\chi^2/\mathrm{ndf} = 8.43/6$ for the high-quality data set. In particular, it shows that the estimated uncertainties of the energy estimator in Sec.~\ref{eventbyeventuncertainties} are compatible with the observed scatter around the calibration curve.

\subsection{Uncertainties of the reconstructed cosmic-ray energy with the radio detector}
To determine the energy resolution of the radio detector, the known resolution of the surface detector needs to be subtracted from the combined scatter. The average (statistical) SD energy resolution for all air showers in our data set is 18\%.  To obtain an estimate of the radio-energy resolution we use a Monte Carlo study which takes into account the energy and zenith angle dependence of the SD energy resolution.  
The combined scatter is simulated for different radio-energy resolutions, according to the number of air showers and the energy and zenith distribution of the data set.  We find that the energy resolution of the radio detector is 22\% for the full data set and 17\% for the air showers where the core position could be determined in the radio LDF fit, when five or more radio stations have a significant signal. 

In the above calculation we assumed that the energy estimates from the SD and radio reconstruction are uncorrelated for a fixed energy. However, an anti-correlation is expected as radio emission originates from the electromagnetic part of the air shower whereas the SD signal is mostly due to muons resulting from the hadronic shower component \cite{ICRC2013Glennys} and which are anti-correlated shower parameters for a fixed cosmic-ray energy. In case of an anti-correlation, the estimated radio-energy resolution would be even smaller making the above values conservative estimates. 

Furthermore, we studied the effect of a possible bias in the SD reconstructed energy for different primaries where the detector is not fully efficient (\unit[0.1]{EeV} - \unit[0.3]{EeV}) and has a slightly different efficiency curve for the two extreme scenarios of proton and iron primaries  \cite{AUGERICRC2011_Maris}. We found that the effect is negligible for our data set. 

The uncertainty on the absolute scale of the energy estimator as discussed in Sec.~\ref{systematics_sradio} is calibrated out by correlating $S_\text{radio}$ with $E_{\text{CR}}$. The method, however, inherits the uncertainties of the SD energy scale. This scale uncertainty is dominated by the FD scale uncertainty, which is used to calibrate the SD. It is 14\% at energies $\geq$ \unit[10$^{18}$]{eV} \cite{EnergyScaleICRC2013} and increases to 16\% at \unit[$10^{17.5}$]{eV}.

\subsection{Precision and possible improvements of the energy reconstruction}

We have found that the instrumental noise and the environmental influences are not the dominant contributions to our energy resolution. Applying the method described to a CoREAS Monte Carlo data set \cite{CORSIKA, CoREAS2013}, including a representative set of shower geometries as well as shower-to-shower fluctuations, but no instrumental or environmental uncertainties, a similar energy resolution is obtained for the same detector layout. 

The intrinsic limitation in the energy resolution due to shower-to-shower fluctuations of the electromagnetic part of the shower is predicted to be smaller than 10\% \cite{Huege2008, LOPES_energyxmax_2014} and we expect that the current energy resolution can be further improved. 
Under the condition that the LDF samples the relevant part of the signal distribution on the ground correctly for all geometries, the energy estimator should only be affected by the shower-to-shower fluctuations in the electromagnetic part of the shower. The only additional geometric dependence is due to the fact that the air shower might not be fully developed when reaching the ground, i.e., some part of the shower is clipped away. As the atmospheric depth increases with the secant of the zenith angle, clipping mostly affects high-energy vertical showers. Hence, we expect an additional dependence on the zenith angle. In the future, with larger statistics, this effect will be parametrized from data and will further improve the energy resolution. 
Also, a better understanding of the detector and the environmental effects, such as temperature dependencies, will help to improve the energy reconstruction.

Combined measurements, such as they are possible at the Pierre Auger Observatory, hold great potential for future improvements of the energy resolution due to the anti-correlation of the energy reconstructed with the radio and surface detectors.

\subsection{The energy content of extensive air showers in the radio frequency range of \unit[30 to 80]{MHz}}

So far, the energy content of extensive air showers in the radio frequency range of \unit[30 to 80]{MHz} has only been measured at the Pierre Auger Observatory in Argentina. However, our findings can be generalized by the following consideration.

To obtain a prediction that is independent of the location of the experiment, i.e., a universal formula to calculate the radiation energy from the cosmic-ray energy, the calibration function Eq. \eqref{eq:calibration_sradio} can be normalized to the local magnetic field. 
We found that it is sufficient to correct only for the dominant geomagnetic part of the radio emission. This is because the increase of radiation energy due to the charge-excess emission is small, as constructive and destructive interference with the geomagnetic emission mostly cancel out in the integration of the energy densities over the shower plane, see Eq.~\eqref{eq:sradio}. For the average relative charge-excess strength of 14\% at AERA \cite{AERAPolarization} the increase in radiation energy is only 2\%. As most locations on Earth have a stronger magnetic field than the AERA site the effect of the charge-excess emission on the radiation energy will be even smaller. Within the statistical accuracy of the calibration function this effect can be neglected which leads to the universal prediction of the radiation energy
\begin{linenomath}\begin{equation}
\begin{split}
 E_\mathrm{\unit[30 - 80]{MHz}} =& \unit[(15.8 \pm 0.7 (\mathrm{stat}) \pm 6.7 (\mathrm{sys}))]{MeV} \\ 
 &\times \left(\sin\alpha \,\frac{E_\mathrm{CR}}{\unit[10^{18}]{eV}} \, \frac{B_\mathrm{Earth}}{\unit[0.24]{G}} \right)^2 ,
\end{split}
\end{equation}\end{linenomath} 
where $E_\mathrm{CR}$ is the cosmic-ray energy, $B_\mathrm{Earth}$ denotes the local magnetic-field strength and \unit[0.24]{G} is the magnetic-field strength at the AERA site. The systematic uncertainty quoted here is the combined uncertainty of $S_\mathrm{radio}$ (28\%) and the SD energy scale (16\% at \unit[$10^{17.5}$]{eV}). This formula will become invalid for radio detectors at high altitudes because the amount of radiation energy decreases as -- depending on the zenith angle -- a significant part of the air shower is clipped away at the ground. 

Please note that in practice the \unit[30 to 80]{MHz} band is used by most experiments. Due to coherence effects, the cosmic-ray-induced radio emission is strongest below \unit[100]{MHz}. Atmospheric noise and short-wave band transmitters make measurements below \unit[30]{MHz} unfeasible. From \unit[85 to 110]{MHz} the FM band interferes with measurements. Furthermore, radio emission at frequencies well beyond \unit[100]{MHz} can be detected only in very specific geometries (observers at the Cherenkov angle) \cite{LOFARCherenkov2015}. Hence, ground-based experiments exploit the frequency window from \unit[30 to 80]{MHz} or measure in only slightly different frequency bands.

\begin{section}{Conclusions}

The Auger Engineering Radio Array is the radio detector of the Pierre Auger Observatory. It is located within the low-energy extension of the Observatory where additional surface detector stations with a smaller spacing are present,
which enables access to cosmic-ray energies down to \unit[0.1]{EeV}.
For the analysis presented here we only use the thoroughly calibrated 24 LPDA radio stations of the first stage of AERA deployment, with data collected between April 2011 and March 2013.

At several observer positions, the energy deposit per area of the radio pulse of an extensive air shower is measured. Using recent progress in understanding the lateral signal distribution of the radio signals, this distribution is described by an empirical function. The spatial integral of the lateral distribution function gives the amount of energy that is transferred from the primary cosmic ray into radio emission in the \hbox{\unit[30 to 80]{MHz}} frequency band of AERA during the air-shower development. We measure on average \unit[15.8]{MeV} of radiation energy for a \unit[1]{EeV} air shower arriving perpendicularly to a geomagnetic field of \unit[0.24]{G}. The systematic uncertainty is 28\% on the radiation energy and 16\% on the cosmic-ray energy.

This radiation energy -- corrected for different emission strengths at different angles between shower axis and geomagnetic field -- is used as the cosmic-ray energy estimator $S_\text{radio}$. A comparison of $S_\text{radio}$ with the cosmic-ray energy of the surface detector reconstruction shows that it is consistent with quadratic scaling with the cosmic-ray energy $S_\text{radio} \propto E^B$ where $B = 1.98 \pm 0.04$ as expected for coherent radio emission.

The calibration function is normalized to the strength of the local geomagnetic field. Hence, with the knowledge of the local geomagnetic field and a measurement of the radiation energy (in the AERA frequency range) the calibration function can be used at any location to calculate the cosmic-ray energy.

Investigating the scatter around the calibration curve and subtracting the resolution of the surface detector we find that the energy resolution of the radio detector is 22\% for the full data set, and 17\% for the events with more than four stations with signal, where the core position could be determined in the radio LDF fit. 
Given the small shower-to-shower fluctuations of the electromagnetic component, we expect that with a deeper understanding of the detector and environmental effects, an even improved precision in the energy measurement can be achieved.

\end{section}

\appendix

\section{LDF parameters}
\label{LDF_parametersr}
Table \ref{tab:c1c2} gives the parameters used in the LDF function of Eq. \eqref{eq:LDF}.
\begin{table}[h]
\caption{Parameters $C_0$ - $C_4$ of Eq.~\eqref{eq:LDF}. $C_3 = \unit[16.25]{m}$ and $C_4 = \unit[0.0079]{m^{-1}}$. The zenith-angle dependent values used to predict the emission pattern are given for zenith angle bins up to $60^{\circ}$. }
\centering
\begin{tabular}{cccc}
\hline \hline
zenith angle&$C_0$&$C_{1} [\unit{m}]$&$C_{2}$ [\unit{m}] \\
\hline
$0^{\circ}-10^{\circ}$&$0.41\,$ &$-8.0\pm 0.3$&$21.2\pm 0.4$ \\
$10^{\circ}-20^{\circ}$&$0.41\,$ &$-10.0\pm 0.4$&$23.1\pm 0.4$\\
$20^{\circ}-30^{\circ}$&$0.41\,$ &$-12.0\pm 0.3$&$25.5\pm 0.3$\\
$30^{\circ}-40^{\circ}$&$0.41\,$ &$-20.0\pm 0.4$&$32.0\pm 0.6$\\
$40^{\circ}-50^{\circ}$&$0.46\,$ &$-25.1\pm 0.9$&$34.5\pm 0.7$\\
$50^{\circ}-60^{\circ}$&$0.71\,$   &$-27.3\pm 1.0$&$9.8\pm 1.5$\\
\hline \hline
\end{tabular}
\label{tab:c1c2}
\end{table}

\section{Likelihood function}
\label{Likelihood}
The likelihood function (for one pair of radio signal $S_\mathrm{radio}$ and SD cosmic-ray energy estimate $E_{\mathrm{SD}}$) has the following form
\begin{linenomath}\begin{eqnarray}
l(S_\mathrm{radio}, E_\mathrm{SD}) =&&\, \frac{1}{\mathrm{N}} \sum\limits_{i}  \frac{\varepsilon_\mathrm{SD}(E_{SD}, \Theta_i) \, \varepsilon_\mathrm{RD}(E_\mathrm{SD}, \Theta_i, \Phi_i)} {\varepsilon_\mathrm{SD}(E_{\mathrm{SD},i}, \Theta_i) \, \varepsilon_\mathrm{RD}(E_{\mathrm{SD},i}, \Theta_i, \Phi_i)}  \nonumber   \\ 
&&\times g_\mathrm{RD}(S_\mathrm{radio} | S(E_{\mathrm{SD},i}), ...) \nonumber \\
&&\times g_{\mathrm{SD-sh}}(E_\mathrm{SD} | E_{\mathrm{SD},i}, \Theta_i) \, .
\end{eqnarray}\end{linenomath}
The summation is performed over all events in the selected data set. $g_\mathrm{RD}(S_\mathrm{radio} | S, ...)$ and $g_\mathrm{SD-sh}(E_\mathrm{SD} | E, \Theta)$ are the conditional probability density functions, which describe the probability to measure a radio signal $S_\mathrm{radio}$ or energy $E_\mathrm{SD}$ if the true radio signal, energy and zenith angle are $S$, $E$ and $\Theta$. $\Phi$ denotes the azimuth angle. $g_\mathrm{RD}(S_\mathrm{radio})$ is obtained for each event in a Monte Carlo simulation where all reconstructed parameters that influence the radio-energy estimator are varied within their uncertainties.
$\varepsilon_\mathrm{SD}(E_\mathrm{SD}, \Theta)$ and $\varepsilon_\mathrm{RD}(E_\mathrm{SD}, \Theta, \Phi)$ are the efficiencies of the surface and the radio detector. 
The radio efficiency has been determined with Monte Carlo air-shower simulations and a full-detector simulation and depends on the energy, the zenith and the azimuth angle.  
$\mathrm{N}$ is the normalization of the function to an integral of one.

\begin{acknowledgments}

The successful installation, commissioning, and operation of the Pierre Auger
Observatory would not have been possible without the strong commitment and
effort from the technical and administrative staff in Malarg\"ue. We are
very grateful to the following agencies and organizations for financial
support:

\begin{sloppypar}
Comisi\'on Nacional de Energ\'{\i}a At\'omica,
Agencia Nacional de Promoci\'on Cient\'{\i}fica y Tecnol\'ogica (ANPCyT),
Consejo Nacional de Investigaciones Cient\'{\i}ficas y T\'ecnicas (CONICET),
Gobierno de la Provincia de Mendoza,
Municipalidad de Malarg\"ue,
NDM Holdings and Valle Las Le\~nas, in gratitude for their continuing cooperation over land access,
Argentina;
the Australian Research Council (DP150101622);
Conselho Nacional de Desenvolvimento Cient\'{\i}fico e Tecnol\'ogico (CNPq), Financiadora de Estudos e Projetos (FINEP),
Funda\c{c}\~ao de Amparo \`a Pesquisa do Estado de Rio de Janeiro (FAPERJ),
S\~ao Paulo Research Foundation (FAPESP) Grants No.\ 2010/07359-6 and No.\ 1999/05404-3,
Minist\'erio de Ci\^encia e Tecnologia (MCT),
Brazil;
Grant No.\ MSMT-CR LG13007, No.\ 7AMB14AR005, and the Czech Science Foundation Grant No.\ 14-17501S,
Czech Republic;
Centre de Calcul IN2P3/CNRS, Centre National de la Recherche Scientifique (CNRS),
Conseil R\'egional Ile-de-France,
D\'epartement Physique Nucl\'eaire et Corpusculaire (PNC-IN2P3/CNRS),
D\'epartement Sciences de l'Univers (SDU-INSU/CNRS),
Institut Lagrange de Paris (ILP) Grant No.\ LABEX ANR-10-LABX-63,
within the Investissements d'Avenir Programme Grant No.\ ANR-11-IDEX-0004-02,
France;
Bundesministerium f\"ur Bildung und Forschung (BMBF),
Deutsche Forschungsgemeinschaft (DFG),
Finanzministerium Baden-W\"urttemberg,
Helmholtz Alliance for Astroparticle Physics (HAP),
Helmholtz-Gemeinschaft Deutscher Forschungszentren (HGF),
Ministerium f\"ur Wissenschaft und Forschung, Nordrhein Westfalen,
Ministerium f\"ur Wissenschaft, Forschung und Kunst, Baden-W\"urttemberg,
Germany;
Istituto Nazionale di Fisica Nucleare (INFN),
Istituto Nazionale di Astrofisica (INAF),
Ministero dell'Istruzione, dell'Universit\'a e della Ricerca (MIUR),
Gran Sasso Center for Astroparticle Physics (CFA),
CETEMPS Center of Excellence, Ministero degli Affari Esteri (MAE),
Italy;
Consejo Nacional de Ciencia y Tecnolog\'{\i}a (CONACYT),
Mexico;
Ministerie van Onderwijs, Cultuur en Wetenschap,
Nederlandse Organisatie voor Wetenschappelijk Onderzoek (NWO),
Stichting voor Fundamenteel Onderzoek der Materie (FOM),
Netherlands;
National Centre for Research and Development, Grants No.\ ERA-NET-ASPERA/01/11 and No.\ ERA-NET-ASPERA/02/11,
National Science Centre, Grants No.\ 2013/08/M/ST9/00322, No.\ 2013/08/M/ST9/00728 and No.\ HARMONIA 5 - 2013/10/M/ST9/00062,
Poland;
Portuguese national funds and FEDER funds within Programa Operacional Factores de Competitividade through Funda\c{c}\~ao para a Ci\^encia e a Tecnologia (COMPETE),
Portugal;
Romanian Authority for Scientific Research ANCS,
CNDI-UEFISCDI partnership projects Grants No.\ 20/2012 and No.\ 194/2012,
Grants No.\ 1/ASPERA2/2012 ERA-NET, No.\ PN-II-RU-PD-2011-3-0145-17 and No.\ PN-II-RU-PD-2011-3-0062,
the Minister of National Education,
Programme Space Technology and Advanced Research (STAR), Grant No.\ 83/2013,
Romania;
Slovenian Research Agency,
Slovenia;
Comunidad de Madrid,
FEDER funds,
Ministerio de Educaci\'on y Ciencia,
Xunta de Galicia,
European Community 7th Framework Program, Grant No.\ FP7-PEOPLE-2012-IEF-328826,
Spain;
Science and Technology Facilities Council,
United Kingdom;
Department of Energy, Contracts No.\ DE-AC02-07CH11359, No.\ DE-FR02-04ER41300, No.\ DE-FG02-99ER41107 and No.\ DE-SC0011689,
National Science Foundation, Grant No.\ 0450696,
The Grainger Foundation,
USA;
NAFOSTED,
Vietnam;
Marie Curie-IRSES/EPLANET,
European Particle Physics Latin American Network,
European Union 7th Framework Program, Grant No.\ PIRSES-2009-GA-246806;
and
UNESCO.
\end{sloppypar}
\end{acknowledgments}

\bibliographystyle{apsrev4-1}
\bibliography{library_ecalpaper}

\end{document}